\input harvmac

\overfullrule=0pt
\def\Title#1#2{\rightline{#1}\ifx\answ\bigans\nopagenumbers\pageno0\vskip1in
\else\pageno1\vskip.8in\fi \centerline{\titlefont #2}\vskip .5in}

\font\ticp=cmcsc10
\font\secfont=cmcsc10

%
%
\baselineskip=18pt plus 2pt minus 2pt

\def\ajou#1&#2(#3){\ \sl#1\bf#2\rm(19#3)}
%
\def\CH{{\cal H}}

%


%


\def\m{\mu}

\def\p{\pi}
\def\e{\epsilon}

\def\l{{\lambda}}

%

\def\[{\left [}
\def\]{\right ]}
\def\({\left (}
\def\){\right )}

\def\TrH#1{ {\raise -.5em
                      \hbox{$\buildrel {\textstyle  {\rm Tr } }\over
{\scriptscriptstyle \CH _ {#1}}$}~}}

\def\IZ{\relax\ifmmode\mathchoice
{\hbox{\cmss Z\kern-.4em Z}}{\hbox{\cmss Z\kern-.4em Z}}
{\lower.9pt\hbox{\cmsss Z\kern-.4em Z}}
{\lower1.2pt\hbox{\cmsss Z\kern-.4em Z}}\else{\cmss Z\kern-.4em Z}\fi}
\def\IC{\relax\hbox{$\inbar\kern-.3em{\rm C}$}}
\def\IR{\relax{\rm I\kern-.18em R}}
\def\1{\relax 1 { \rm \kern-.35em I}}
\font\cmss=cmss10 \font\cmsss=cmss10 at 7pt

%

\def\frac#1#2{{#1 \over #2}}
\def\ie{{\it i.e.}}

\def\p+{{\partial_+}}

\def\apm{\alpha^{\prime}}


%

\Title{\vbox{\baselineskip12pt
\hbox{\ticp TIFR/TH/97-02}
\hbox{hep-th/9702050}
}}
{\vbox{\centerline {\bf MICROSTATES OF NON-SUPERSYMMETRIC
BLACK HOLES} }}

\centerline{{\ticp
Atish Dabholkar\footnote{$^\dagger$}{e-mail: atish@theory.tifr.res.in}
}}

\vskip.1in
\centerline{\it Tata Institute of Fundamental  Research}
\centerline{\it Homi Bhabha Road, Mumbai, India 400005.}

\vskip .1in

\bigskip
\centerline{ABSTRACT}
\medskip

A five-dimensional dyonic black hole in Type-I theory is
considered that is extremal but non-supersymmetric. It is
shown that the Bekenstein-Hawking entropy of this black
hole counts precisely  the microstates of a D-brane
configuration with the same charges and mass, even though
there is no apparent supersymmetric nonrenormalization
theorem for the mass. A similar result is known  for the
entropy at the stretched horizon of electrically charged, extremal,
but non-supersymmetric black holes in heterotic string theory.
It is argued that  classical nonrenormalization
of the mass may partially explain this result.

\bigskip

\bigskip
\Date{February, 1997}

\vfill\eject

\def\npb#1#2#3{{\sl Nucl. Phys.} {\bf B#1} (#2) #3}
\def\plb#1#2#3{{\sl Phys. Lett.} {\bf B#1} (#2) #3}
\def\prl#1#2#3{{\sl Phys. Rev. Lett. }{\bf #1} (#2) #3}
\def\prd#1#2#3{{\sl Phys. Rev. }{\bf D#1} (#2) #3}

\def\mpl#1#2#3{{\sl Mod. Phys. Lett. }{\bf #1} (#2) #3}

%
\lref\WiOl{E. Witten and D. Olive, \plb{78}{1978}{97}.}

\lref\Wittbound{E. Witten, \npb{460}{1996}{335},
{hep-th/9510135}.}

\lref\Wittsmall{E. Witten,  \npb{460 }{1996}{541}, {hep-th/9511030}.}

\lref\Wittdual{E. Witten, \npb{443}{1995}{85}, hep-th/9503124.}

\lref\DaHa{ A. Dabholkar and J. A. Harvey,
\prl{63}{1989}{478}.}

\lref\DGHR{ A. Dabholkar, G. Gibbons, J. A. Harvey, and F. R. Ruiz-Ruiz,
\hfill\break \npb{340}{1990}{33}.}

\lref\DGHW{A. Dabholkar, J.~P.~Gauntlett, J.~A.~Harvey, and D.~Waldram,
\hfill\break \npb{474}{1996}{85}, hep-th/9511053.}

\lref\Dabh{A. Dabholkar,
\plb{357}{1995}{307}, hep-th/9506160.}

\lref\Hull{C. M. Hull, \plb{357}{1995}{345}, hetp-th/9506194.}

\lref\CvYo{ M. Cvetic and D. Youm, { hep-th/9510098}.}

\lref\Tsey{A. Tseytlin, \mpl{A11}{1996}689, {hep-th/9601177}.}

\lref\CvTs{ M. Cvetic and A. Tseytlin, \prd{53}{1996}{5619},
{hep-th/9512031}.}

\lref\LaLi{L. D. Landau and E. M. Lifshitz,
{\it The Classical Theory of Fields}, 4th edition, Pergamon Press (1989),
$\S{96}$.}

\lref\Vafa{C. Vafa, \npb{463}{1996}{435}, { hep-th/9512078}.}

\lref\StVa{A. Strominger and C. Vafa, \plb{379}{1996}{99},
{hep-th/9601029}.}

\lref\GHS{D. Garfinkle, G. Horowitz and A. Strominger,
\prd{43}{1991}{3140}.}

\lref\HoSt{G. Horowitz and A. Strominger, \prl{77}{1996}{2368},
hep-th/9602051.}

\lref\HMS{G. Horowitz, J. Maldaena, A. Strominger, \plb{383}{1996}{151},
hep-th/9603109.}

\lref\KLMS{D. M. Kaplan, D. A. Lowe, J. M. Maldacena, and A. Strominger,
hep-th/9609204}

\lref\LaWi{F.~Larsen and F.~Wilczek, \plb{375}{1996}{37}, hep-th/9511064.}

\lref\Duff{M. Duff, J. T. Liu, and J. Rahmfeld, hep-th/9612015.}

\lref\DuRa{M. Duff and J. Rahmfeld, \npb{481}{1996}{332}, hep-th/9605085.}

\lref\KhOr{R. Khuri and T. Ortin, \plb{373}{56}{1996}, hep-th/9512178\semi
T. Ortin, hep-th/9612142.}

\lref\Polc{J. Polchinski, \prl{75}{1995}{4724}.}

\lref\Dbrane{J.~Dai, R.~Leigh, and J.~Polchinski, \mpl{A4}{1989}{2073}\semi
P.~Horava, \plb{231}{1989}{251}.}

\lref\PoWi{J.~Polchinski and E.~Witten, \npb{460}{1996}{525},
hep-th/9510169.}

\lref\GiPo{E.~Gimon and J.~Polchinski, \prd{54}{1996}{1667}, hep-th/9601038.}

\lref\Senentropy{A. Sen, {\sl Mod. Phys. Lett.} {\bf A10} (1995) 2081,
hep-th/9504147.}

\lref\Senblack{A. Sen, \npb{440}{1995}{421}, hep-th/9411187.}

\lref\Senfour{A. Sen, {\sl Int. J. Mod. Phys.}
{\bf A9}  (1994) 3707,  hep-th/9402002.}

\lref\CaMa{C. G. Callan and J. M. Maldacena,  \npb{475}{1996}{645},
hep-th/9602043.}

\lref\CMP{C. G. Callan, J. M. Maldacena, and A. W. Peet, \npb{475}{1996}{631},
hep-th/9510134.}

\lref\Peet{A. W. Peet,  \npb{446}{1995}{211}, hep-th/9504097.}

\lref\HuPo{J. Huges and J. Polchinski, \npb{278}{1986}{147}.}

\lref\DMR{A. Dabholkar, G. Mandal, and P. Ramadevi, {\it work in progress}}

\lref\HLM{G. Horowitz, D. Lowe, and J. Maldacena, \prl{77}{1996}{430},
hep-th/9603195.}

\newsec{Introduction}

The spectrum of supersymmetric states  has played a crucial
role in understanding strong coupling phenomena  in string theory.
A supersymmetric state (or a `superstate' for short) is a state
that preserves some of the supersymmetries, and
belongs to a short representation of the supersymmetry algebra \WiOl.
An important
property of a superstate that follows from the supersymmetry algebra is that it
is always extremal, \ie, its mass $M$ is always
proportional to the absolute value of some charge $Z$. This also
implies that a non-extremal state is always non-super, but the converse
is not always true. In fact, we shall be interested in
precisely such states, in Type-I and heterotic string theory, that
are extremal but non-super\foot{In recent
literature a superstate is commonly called a BPS state. However,
before the work of Witten and Olive \WiOl,
a BPS state meant a state that was only classically extremal.
To distinguish between the two notions,
we have used the terms `super' and `extremal'
instead of `BPS'.}.

The significance of superstates to strong coupling physics stems
from the fact that their exact spectrum can often be computed
reliably in the semiclassical regime. One expects that, under
suitable conditions, the
number of states cannot jump discontinuously as we smoothly
vary the coupling constant. In particular,
a short multiplet at weak coupling is expected to
remain short even at strong coupling. Moreover,
because the extremality of a short multiplet is a consequence
of the supersymmetry algebra, its mass is proportional to the charge
even at strong coupling. With enough supersymmetries, the charge
is sometimes not renormalized, which for the superstates
implies that the  mass is also not renormalized.
This property of exact extremality is crucial for the stability
of the superspectrum. Charge conservation together with
energetic considerations are enough to ensure the
stability of many superstates.

These special  properties of the superspectrum have
proved to be extremely useful  recently, in particular, for obtaining
a statistical interpretation in terms of the underlying microstates
of the Bekenstein-Hawking
entropy of certain  supersymmetric black holes in string theory.
In string theory, the spectrum of superstates is much richer
than in field theory. For a given mass and charges, there
is usually a large degeneracy of superstates which can be
counted reliably at weak  coupling.
As one increases the strength of the coupling,  the superstate
eventually undergoes a gravitational collapse, and forms a black hole.
Because the spectrum of the superstates does not change as
we vary the coupling, the  degeneracy of the corresponding black-hole
states at strong coupling must be  the same as the degeneracy
of superstates at weak coupling. This degeneracy
has been shown to be in precise numerical agreement
with the exponential of the Bekenstein-Hawking entropy in a large
number of cases.
In this paper, we shall be interested in a similar counting of  states for
a certain class of non-supersymmetric but extremal black holes
that exist in toroidally compactified heterotic and Type-I string theory.
In this case,  supersymmetry alone does not protect the spectrum
and there is no {\it a priori} reason to expect that the degeneracy at
strong coupling should be the same as that  at weak coupling. Surprisingly
we still find this to be the case.

Our discussion of  the non-supersymmetric states and the associated
black holes will closely parallel that of the supersymmetric states.
So, let us begin by recalling some facts about the superstates.
One simple class of
superstates are the perturbative, electric superstates in
toroidally compactified
heterotic string theory \DaHa.
Consider heterotic string compactified on an
$n$-dimensional torus to $D=10-n$ dimensions.
There are
$16$ real supersymmetries corresponding to the
 $16$ real components of a single Majorana-Weyl fermion
in ten dimensions. All supersymmetries are carried by
the right-movers.  To discuss the spectrum it is convenient to use the
Green-Schwarz formalism in the light-cone gauge. For a state
of mass $M$ and charges $q_L$ and $q_R$ the Virasoro constraints
are
\eqn\hetcon{
   N_L =1 + \apm\left({1\over 4}M^2- q_L^2\right), \qquad
   N_R =\apm\left({1\over 4} M^2 -q_R^2\right),}
where $N_L$, and $N_R$ are  the number of transverse left-moving and
right-moving  oscillators respectively.
A short representation corresponding to a superstate
in spacetime that preserves $8$ of the original $16$ supersymmetries
can be obtained only
if all right-moving oscillators on the world-sheet are in the  ground state,
\ie\ $N_R=0$. The mass-shell conditions \hetcon\ can
then  be written as:
\eqn\massright{
M^2=4q_R^2, \qquad
N_L=1 + \apm (q_R^2 -q_L^2).}
For large $N_L$,  the degeneracy of these superstates $d(N_L)$
goes as $d(N_L) \sim e^{4\pi\sqrt{N_L}}$ .
At large coupling, the superstate is described by
a supersymmetric black hole.  For these black holes the event horizon
and the singularity coincide; so at first sight,
the area $A$ of the event horizon and
consequently the Bekenstein-Hawing  entropy $S =A/4$ appears to vanish.
However, Sen \Senentropy\  was able to show  that, in four dimensions,
if one computes the area at the stretched horizon,
which is roughly string-length away from
the event horizon,  then the entropy
is indeed proportional to the logarithm of the
degeneracy of the perturbative superstates.
Similar result holds for higher dimensions as well \Peet.

Instead of considering the superstates that have $N_R=0$,
let us now consider states that have $N_L=1$, but arbitrary $N_R$.
The mass-shell condition now becomes
\eqn\massleft{
M^2=4q_L^2,\qquad
N_R= -\apm (q_R^2 -q_L^2).
}
These states are still extremal at the classical level because
the mass is proportional to $2|q_L|$, but they are no longer super
because, for $N_R\neq0$, they break all the right-moving supersymmetries,
and belong to a long representation.
At tree level,  the perturbative states still have
 the Hagedorn degeneracy that goes as $d(N_R) \sim e^{\sqrt{N_R}}$.
However, now supersymmetry
no longer protects the  mass formula, and in general the mass would
be renormalized. Therefore, the degeneracy of states at strong
coupling can be very different from this classical, perturbative formula.
On the other hand, we can calculate the entropy at the stretched
horizon for the corresponding extremal black holes as suggested
in \refs{\Duff, \DuRa}.
The calculation is identical to that of Sen for the supersymmetric
black holes because it depends only
on the low-energy bosonic fields which are insensitive to the
orientation of the string apart from the labeling of the gauge fields.
{}From \Senentropy, we see that
in the normalizations of
\Senentropy, the  entropy is given by
\eqn\entropyone{
S_{B-H} = \frac{A}{4G_N} \sim  \frac{2\pi C}{g} \sqrt{M^2 -\frac{Q_R^2}{8g^2}}
\ , }
where $m$ is the mass of the black hole in the Einstein metric, $ Q_L$ is the
charge, $g$ is the string coupling constant, and $ C$ is a numerical constant.
The statistical entropy calculated from the number of superstates is given by
\eqn\hagentropy{
S_{stat} = \log{d(N_R)} \sim  \sqrt{N_R} \sim \frac{1}{g}
 \sqrt{M^2 -\frac{Q_R^2}{8g^2}}\ ,}
after converting $ M$ in \massleft\ which is measured in the string metric
and the charges $ q_L$ to Sen's normalizations.
Thus, the  Bekenstein-Hawking entropy agrees, modulo a numerical
coefficient, with the perturbative counting of states.
Now,  even though the bosonic fields are insensitive to
the difference between left-moving and right-moving charges and oscillations,
the fermionic fields and the supersymmetry transformations do notice
this difference \DGHW.
In particular, black hole solutions with $N_R =0$  preserve half the
supersymmetries,
whereas the black hole solutions with $ N_L =1$ but $ N_R \neq 0$
break all supersymmetries.
We thus seem to obtain a surprising  agreement between the Bekenstein-Hawking
entropy and the perturbative counting of states for extremal but
nonsupersymmetric black holes \refs{\Duff, \DuRa}.

One of the disadvantages
of purely electric extremal black holes is the need to consider the stretched
horizon, which introduces the undetermined
numerical coefficient $ C$ in \hagentropy.
In the supersymmetric case,  this problem was
overcome by considering dyonic black holes that
have regular event horizon with nonzero area \StVa.
The Bekenstein-Hawking entropy is then
simply  given by  a quarter of the horizon area
which agrees precisely, without any undetermined
numerical constant, with the  microscopic counting of states.
In this  case, the microscopic superstates at weak coupling
are not perturbative,
but are given by a configuration of D-branes \refs{\Polc, \Dbrane}.
The degeneracy of  these states can be calculated
at weak coupling using D-brane techniques.

Our objective will be to find non-supersymmetric but
extremal black holes that also have a regular event horizon
and nonzero area and compare  it with the D-brane counting.
Consider Type-I theory on $M^5 \times S^1 \times T^4$
where  $ M^5$ is the
five-dimensional Minkowski spacetime with co-ordinates
 $(t, x_1, x_2, x_3, x_4)$,  $S^1$ is a circle
of radius $R$ with co-ordinate $ x_5 \equiv x_5 + 2\pi R$, and
$T^4$ is a four-torus
with co-ordinates $(x_6, x_7, x_8, x_9)$. Now, consider a configuration
of $Q_1$ Dirichlet 1-branes wrapped around the circle,
$Q_5$ Dirichlet 5-branes wrapped around the torus $ S^1 \times T^4$,
and $n$ units of right-moving quantized momentum flowing along the circle.
This particular configuration is motivated by the following
reason.  Consider the
$ SO(32)$ heterotic string compactified on a circle of radius $R$,
and consider states that have winding number $m$ and momentum
$ n/R$ along the circle but no $ SO(32)$ charges.
For these states, we can take $q_L$ and $ q_R$ appearing in \massleft\
and \massright\ to be
\eqn\momenta{q_L = (\frac{n}{2R} + {m R\over 2\apm}),
    \qquad q_R =(\frac{n}{2R} - {m R\over 2\apm}).}
One can then consider either the superstates for which
$ N_R =0$ and $N_L=1 -mn$, or the non-super but extremal
states for which $ N_L=1$ and $ N_R = mn$. The important
difference between  the two cases is that because $ N_R$ and
$ N_L$ are always non-negative,
$ mn$ is negative for the super-states but positive for
the non-super-states.  Thus,  for a  string that is wound
with a positive orientation
($ m >0$), the superstates have left-moving momentum
whereas the non-superstates have right-moving momentum.
Under the duality between heterotic and Type-I theory
\Wittdual, a fundamental, supersymmetric, winding string without oscillations
is mapped
onto the heterotic soliton string \refs{\Dabh, \Hull} which at weak
coupling is described by the Dirichlet heterotic 1-brane \PoWi.
An oscillating non-supersymmetric but extremal
heterotic solitonic solution carrying right-moving momentum
along the string with  large and nonzero $ N_R$ can be obtained as in
\refs{\DGHW, \CMP}.  Dimensional reduction of the oscillating solitonic
string gives an electric, extremal black hole with vanishing horizon area.
If we want a solution  with nonzero area,  we  can add
5-branes. We are thus led to the configuration discussed above.

The dyonic state in Type-I theory couples only to the graviton $ G_{MN}$
and the dilaton $ \phi$from the NS-NS sector, and the 2-form potential
$ B_{MN}$ from the R-R sector. The low energy action for these fields
is
\eqn\action{S = {1 \over {16 \pi G_{10}}}\int d^{10}x \sqrt{-G} \left(
                e^{-2 \phi}(R + 4(\nabla \phi)^2)
                 - {1 \over12} H^2 \right).}
The solution with three charges $Q_1$,  $ Q_5$, and $ n$
is the same as the one that has been considered in Type-IIB
theory before \refs{\Tsey, \CvTs, \CvYo, \LaWi, \HoSt, \CaMa}.
In the normalizations of {\HMS} we have,
\eqn\dil{ H= 2 r_5^2 \e_3 + 2 r_1^2  e^{-2 \phi}  *_6 \e_3,\qquad
e^{-2\phi } = \(1+   { r_5^2\over r^2 }\)
\(1 + {r_1^2 \over  r^2 }\)^{-1},\quad}
\eqn\metric{\eqalign{
dS^2 = &
 \( 1 + { r_1^2 \over r^2}\)^{-1/2} \( 1 + { r_5^2 \over r^2}\)^{-1/2}
 \[ - dt^2 +dx_5^2 +
{r_n^2  \over r^2} (dt -  dx_5)^2
 +\( 1 + { r_1^2  \over  r^2}\) dx_i dx^i \] \cr
 +& \( 1 + { r_1^2  \over r^2}\)^{1/2}\(
1 + { r_5^2 \over r^2}\)^{1/2} \left[
dr^2 + r^2 d \Omega_3^2 \right]
,}}
where $*_6$ is the Hodge dual in the six dimensions $x_0,..,x_5$,
$\e_3$ is the volume element on the unit three-sphere, and $x^i$,  $i =
6,...,9$,
are the co-ordinates of the torus with volume $ {\left( 2\pi \right)^4 V}$.
This solution represents a black hole in the non-compact five dimensions.
The parameters of the solution are related to the integral (positive) charges
through the relations
\eqn\paramsol{
r_1^2 = {\l Q_1 \apm \over V }, \qquad
r_5^2 = {\l Q_5 \apm },\qquad r_0^2
= { \l^2 n \apm \over R^2 V },\qquad
}
where $ \l$ is the string coupling constant.
If we take $ r_5$ to be zero in this solution, then we simply obtain
the heterotic solitonic string carrying momentum \DGHW\ but
no oscillations.
The supersymmetry of such a soliton has been discussed
in detail in \DGHW. We only summarize the conclusions here.
The term  involving $(dt -  dx_5)^2$   in the metric
\metric\ corresponds to right-moving momentum along the soliton
($ mn >0$), and, as expected, the solution breaks all supersymmetries.
If we replace this combination by  $(dt + dx_5)^2$ instead, then the soliton
carries
left-moving momentum and  preserves half the supersymmetries.
Now, if we start with the nonsupersymmetric solution and add 5-branes,
all supersymmetries still remain broken giving us the configuration
we are interested in.

The Bekenstein-Hawking entropy and the energy of this
black hole can be easily computed \refs{\HoSt, \CaMa}:
\eqn\bhentropy{
S_{B-H}= { A_{10} \over 4 G_{10} } =
2\pi\sqrt{Q_1 Q_5 n}, \qquad E ={R Q_1\over \l} +{R V Q_5\over \l} + {n\over
R}.}
These expressions are the same for solutions carrying either
right-moving or left-moving momentum. In particular, both
non-supersymmetric and supersymmetric solutions
are extremal.

We would now like to see if the counting of D-brane states at weak
coupling can reproduce this entropy. In Type-I theory there are
$ 32$ 9-branes in addition to the 1-branes and the 5-branes. The open
strings can end on any of these different type of branes. The
$ (1, 1)$ strings with both ends on the 1-brane gives rise to the
right-moving superstring, whereas the $ (1, 9)$, and $ (9,1)$ strings
give rise to left-moving current algebra of $ Spin(32)/{\IZ_2}$ in the
fermionic
representation. These sectors are thus chiral.
The sectors that are most relevant for the entropy counting
are the $(1, 5)$ and the $(5, 1)$ strings. Recall that the Type-I theory
is an orientifold of Type-IIB theory with orientifold projection
${1 +\Omega \over 2}$ where $ \Omega$ is the worldsheet parity.
Under $ \Omega$ the $(1, 5)$ sector is identified with  the $ (5, 1)$
sector which halves the number of states compared to Type-IIB theory.
On the other hand,
the 5-brane of  Type-I with unit magnetic charge is really two 5-branes
of Type-IIB put together \refs{\Wittsmall, \GiPo}, which doubles the number
of states. So in the end we have effectively
$ 4Q_1Q_5$ superconformal free fields describing a superconformal
sigma model on $ (T^4)^{Q_1Q_5}/S(Q_1Q_5)$ exactly as
in the Type-IIB case \StVa. The counting
of states is the same as that of right-moving oscillators at level $n$
in a conformal field theory with central charge $ c ={3\over 2} (4Q_1 Q_5)$,
which is given by $ d(n) = e^{2\pi \sqrt {c n/6}}$. The statistical entropy
is then
\eqn\statentropy{
S_{stat}= \log{d(n)} = 2\pi\sqrt{Q_1 Q_5 n}\ , }
surprisingly, in precise agreement with \bhentropy.

The lack of supersymmetry of the black-hole solution when the
momentum is right-moving can be seen in the D-brane picture
as well. The spacetime supersymmetries are only partially broken
by a D-brane in the  ground state and the unbroken supersymmetries
give rise to the supersymmetries of the worldvolume theory \refs{\HuPo}.
The  heterotic solitonic string is chiral and the unbroken spacetime
supersymmetries give rise to the supersymmetry only in the right-moving
sector.  The ground state preserves these worldvolume supersymmetry
but a state with nonzero right-moving energy breaks it completely.
Consequently,  for these states,
the spacetime supersymmetry is also completely broken.

We should emphasize here that these black holes are not
nearly supersymmetric but are far from being supersymmetric.
There is no  small parameter that measures the
deviation from supersymmetry.
The entropy of  extremal and nonsupersymmetric
black holes has been considered earlier in a different
context \HLM\ of
rotating black holes in type-II theory in four and
five dimensions, where a similar  agreement was
found with D-brane counting.
In those examples it is the nonzero rotation that breaks supersymmetry
completely even with  extremality. Another example in four
dimensions that has been considered before can be found in
\KLMS. General black hole solutions in arbitrary dimensions
that are supersymmetric but not extremal can be found in
\KhOr.

We do not fully understand how to explain
this  agreement between the counting of
states at weak and strong coupling for these non-supersymmetric states.
We present the following observations which may be relevant.
The coupling constant
of the theory and therefore the charges of these states
are not renormalized by supersymmetry.
It may be that the mass of these states is not renormalized
for reasons other than supersymmetry.  If this is true,
then the
states that are tagged by a particular mass and charges
at weak coupling
would continue to have the same mass and charges even at
strong coupling, and then the counting can agree.
It does appear that for the electric, perturbative states
in heterotic string theory given by  both \massright and \massleft, the mass
is indeed not renormalized classically.
Let us first see how one might
define the notion of classical renormalization.
There are
three massless fields  $ G_{MN}$, $ \phi$, and $ B_{MN}$ that
couple to the states. The classical renormalization can
be defined as the sum of self-energies in these three fields.
For the graviton there is no covariant stress-tensor but one
can define the Landau-Lifshitz energy-momentum pseudo-tensor
$ t_{LL}^{MN}$ \LaLi. The total conserved
stress-tensor $ \Theta^{MN}$ is then the sum
of the stress-tensor for matter $ T^{MN}$ and $ t_{LL}^{MN}$.
For a given metric that satisfies the
Einstein equation,  the total stress-tensor is
\eqn\emtensor{
\Theta^{PQ}= -{1 \over 16\pi G_{D}} \partial_M \partial_N
\left[ g \left( g^{PQ} g^{MN} - g^{PM} g^{QN} \right) \right],
}
where $ g_{MN}$ is the Einstein metric with determinant $ g$,
and $G_{D}$ is Newton's
constant in  $ D$ dimensions. We take spacetime to be a product
of $\left( D-1 \right)$ dimensional Minkowski spacetime and a circle of radius
$ R$. The radius can be quite small but still larger
than the string scale.
Let us now consider an oscillating  heterotic string that wraps around
this circle with {\it right-moving} oscillations.  The solution for the
massless fields describing  such
a state has been discussed in \refs{\DGHW, \CMP}.
Let $ \left( t, z \right)$ be the longitudinal coordinates and $ y^i$, $
i=1,...,(D-2)$
the transverse coordinates.
The dilaton $ \phi$ satisfies the linear equation
$ \partial^2 \left( e^{-2\phi} \right) = -16\pi \m G_D \delta^{D-2} (y) $
where $ \partial^2$ is the flat space Laplacian of the transverse
co-ordinates, and $ \m$ is the bare string tension.
The line-element in  the Einstein metric is
\eqn\einmetric{
ds^2 = e^A \left[-dt^2 + dz^2 + \vec{f} \cdot \vec{y} (dt - dz)^2
\right]
+e^B\left( dy^i dy^i \right),
}
where $ A =2\left( D-4 \right)\phi/(D-2)$, $ B=-4\phi/(D-2)$,
and $ \vec{f}$ is the profile of oscillation that depends only
on $ (t-z)$. For this time-dependent solution, all quantities like
mass will, in general, be time-dependent.
Here we consider the  time-averaged mass.
Substituting the metric in \emtensor, and time-averaging
over a period of oscillation, we obtain the
total energy-density:
\eqn\edensity{
\rho_{total} = -{1\over 16\pi G_{D}} \partial^2 \left( e^{-2\phi} \right) =
\m\delta^{D-2} (y).
}
It is striking that the self-energy contributions due to
various massless fields cancel precisely leaving only
a delta-function source with
the bare string tension $ \m$. The mass of these nonsupersymmetric
but extremal electric states is therefore not renormalized classically
if we consider only the massless fields.
The massive fields can presumably be included in this analysis
by adding the $ \apm$ corrections to the low-energy action for
the massless fields.

Actually, for our purpose we do not really need a pointwise cancellation
of self-energies as above. All that is needed is that the ADM mass be
the same as the mass of a string source at the origin. For this
purpose we need to match the solution onto string sources at the origin
and then go to a coordinate system that is asymptotically flat.
For the superstates \massleft\ with {\it left-moving} oscillations, this
is described in detail in $\S 2.6$ of \DGHW,
where it is shown
that indeed there is no classical renormalization of the mass.
Now,  even for the non-super states, the classical calculation is the same.
So we conclude that classically the mass of the extremal non-super
states is also not renormalized.
For the supersymmetric
states,
the classical nonrenormalization is a reflection
of the nonrenormalization at the quantum level \refs{\DaHa, \DGHR}.
It is very interesting to know, therefore, whether for the
non-supersymmetric states also,  the classical
nonrenormalization of mass
continues to hold at the quantum level \DMR.
It is tempting to speculate that there
may be some hidden gauge invariance which can explain this
nonrenormalization even without supersymmetry.

One final remark is that
for a {\it dyonic} black hole with finite horizon area,
or for the electric solutions carrying a  pure momentum wave
we do not know of an analogous result.
Recall that a pure momentum wave without oscillations is obtained
by replacing
$\vec{f} \cdot \vec{y} $ in \einmetric\ by $ p/y^{D-4}$, but
then the solution
does not match onto a  classical delta-function source \DGHW.
In these cases, however,  when there is  an event-horizon, or when the solution
does  not match onto classical sources, it is not clear how
to define the notion of bare mass or  self-energy in a meaningful way
to begin with.

\bigskip
\leftline{ \secfont Acknowledgements}
\bigskip

I would like to thank Mike Duff, Ashoke Sen, and Lenny Susskind
for useful discussions, Gary Horowitz, Jeff Harvey, and Juan Maldacena
for comments on the draft, and the organizers of the
1996 Puri Workshop for inviting me to a very stimulating
meeting where some of this work was done.
\vfill
\eject
\listrefs
\end